\def\be{\begin{equation}}
\def\te{\end{equation}}
\def\bea{\begin{eqnarray}}
\def\nn{\nonumber}
\def\tea{\end{eqnarray}}

\documentstyle[12pt]{article}
\makeatletter 
\@addtoreset{equation}{section}
\makeatother  

\begin{document}

\title{Correlations, Decoherence, Dissipation, and Noise
in Quantum Field Theory
\thanks{Invited talks given by the authors at the International Workshop on
``Heat Kernel Techniques and Quantum Gravity", August 1994, Winnepeg, Canada.
Proceedings in  {\it Discourses in Mathematics and
Its Applications, No.~4} edited by S. A. Fulling
(Texas A $\&$ M University Press, College Station, 1995)} }
\author{Esteban  Calzetta\\
{\small IAFE and FCEN,  Buenos Aires, Argentina}\\
B.  L.  Hu\\
{\small Department of Physics, University of Maryland, College Park, MD
20742}\\
{\small Institute for Advanced Study, Princeton, New Jersey 08540}}
\date{\today}
\maketitle
\centerline{(IASSNS-HEP-95/2, umdpp 95-079)}
\begin{abstract}
The statistical mechanical properties of
interacting quantum fields in terms of the dynamics of the correlation
functions are investigated.
We show how the Dyson - Schwinger equations may be derived from a
formal action functional, the n-particle irreducible
($nPI, n \rightarrow \infty$) or the `master' effective action.
It is related to the decoherence functional between histories defined in
terms of correlations. Upon truncation of the Dyson - Schwinger hierarchy at
a certain order, the master effective action becomes complex,
its imaginary part arising from the higher order correlation functions,
the fluctuations of which we define as the correlation noises of that order.
Decoherence of correlation histories via these noises gives rise to
classical stochastic histories
Ordinary quantum field theory corresponds to taking
the lowest order functions, usually the mean field and the 2-point functions.
As such, our reasoning shows that it is an effective theory which can be
intrinsically dissipative. The relation of loop expansion and correlation
order as well as the introduction of an arrow of time from the choice of
boundary conditions are expounded with regard to the origin of dissipation
in quantum fields. Relation with  critical phenomena, quantum transport,
molecular hydrodynamics and potential applications to
quantum gravity, early universe processes and black hole physics are mentioned.

\end{abstract}

\newpage

\section{Introduction and Summary}

\subsection{Background}

The present work is a continuation of our systematic investigation
since 1986 on the statistical and stochastic natures of quantum
field theory. It has developed in three stages incorporating three key ideas.
Our 1988 work \cite{ch88} on non-equilibrium
quantum fields studied the kinetic theory aspects of interacting quantum
field theory. The closed time path (CTP) \cite{ctp}
n-particle irreducible (nPI)
\cite{2pi} effective action for the correlation functions was applied to the
example of a $\lambda \phi^4$ theory to derive the quantum field
equivalent of the BBGKY hierarchy of equations. We derived the relativistic
quantum field Boltzmann equation and explained the origin of dissipation
in a three-loop perturbation analysis. In there, the first key concept of
{\it correlation and dissipation} in the Boltzmann context
was expounded.

Investigating the missing link with noise and fluctuations which
are expected to accompany  dissipation \cite{HuPhysica} led to the
second stage of development which involved
the application of quantum open system concepts \cite{qos}
and influence functional (IF) techniques \cite{if} to quantum Brownian models
(QBM) \cite{qbm} and quantum field theory \cite{HM2,HM3}.
There it was found that the effect  of an environment on the system
can be best depicted by an influence action which contains a dissipative
kernel and a noise kernel as its real and imaginary parts, the two
always balanced by a fluctuation-dissipation relation \cite{fdt}.
The noise is generally colored, being determined by the spectral density of
the environment and the type of coupling between the system and the
environment.
The influence functional formalism makes it possible to derive from first
principles the nature of quantum noise in terms of the fluctuations of
quantum fields \cite{nfsg,fdrsc} and a Langevin or
Fokker-Planck equation for the dissipative dynamics of the system
driven by such stochastic sources. In there, the second key concept of
{\it noise and fluctuations} and their relation
with dissipation are explored  in the Langevin context.

The third key concept
is {\it decoherence}, which predicates the transition of a system from quantum
to classical. In the environment-induced decoherence scheme \cite{envdec}
its effect  can be seen as a direct consequence of the system's
coupling to an environment and the coarse-graining of the environment.
The formulation in terms of consistent histories
\cite{conhis} is an important conceptual advance
in the interpretation of quantum mechanics. These programs in
decoherence relate decoherence in the system to  noise and
fluctuations in the environment.
Applying the consistent history scheme
to correlations of histories, we formulated a new way to describe
decoherence and quantum to classical transition for a closed system, where
the coarse-graining is determined by the correlation order. It is, in our
opinion, more natural than the open system decoherence schemes
where the system- environment separation and the coarse-graining  (by
averaging out some variables in the environment or by the use of smearing or
gate functions in path-integrals) are put in by hand.
Closing this line of development of ideas, we further showed \cite{nfsg} that
the closed time path formulation of field theory is largely equivalent to the
influence functional formulation from which one can easily derive
the Langevin equation driven by a stochastic source (noise)
related to the fluctuations of the quantum fields
for the dissipative dynamics of an open system. Our present work is a further
development of our previous work \cite{ch88,DCH,nfsg},
linking  up the three pairs of key relations amongst
{\it correlation, decoherence, dissipation and noise}. The interconnection
of these effects has been emphasized before by Hu \cite{HuTsukuba} and
Gell-Mann and Hartle \cite{GelHar2}.

As background, some key ideas described above have been explored in the context
of quantum mechanics \cite{UncPri}, gravitation and cosmology
\cite{HuRevs,pazsinha,HuBelgium,Banff}.
On a grander scale, our program here strives to establish
a statistical mechanics of quantum fields. As we shall see,
the physical meaning of
many common techniques and approximation schemes in perturbative field theory
can be truly understood only by bringing in some basic concepts and techniques
of statistical mechanics \cite{Landau,Huang,Itzykson}, kinetic theory
\cite{Akhiezer,Balescu,deGroot}, stochastic mechanics \cite{vanKampen,Weiss},
critical dynamics \cite{Ma,Zinn},
and hydrodynamics \cite{BoonYip,Spohn}.
For example, the construction of the effective
action \cite{effaction}, which describes the dynamics of the mean field,
falls under the general pattern of identification of a subdynamics
\cite{projop} through coarse graining and truncation of a more complete
underlying theory. The meaningfulness of such a division depends on the
identifiability and viability of different space, time and interaction scales
of the characteristic physical processes involved \cite{vanKamp}.
Of course, this is just one instance of the kind of
process by which many body theory and kinetic theory are reduced to
hydrodynamics \cite{Spohn}, or, for that matter, any
microscopic description of a system is coarse-grained into some macroscopic
dynamics of its collective modes \cite{Balian}.

\subsection{Summary}

We shall now summarize the main themes and key findings of this paper.

\subsubsection{ Main Themes}

In this paper we start from the thesis that the full dynamics of an interacting
quantum field may be described by means of the Dyson- Schwinger equations
governing the infinite hierarchy of Wightman functions which measure the
correlations of the field. We show how
this hierarchy of equations can be obtained from the variation of the infinite
particle irreducible, or `master' effective action. Truncation of this
hierarchy
gives rise to a quantum subdynamics governing a finite number of the
correlation
functions (which constitute the `system') and expression of the higher
order correlation functions (which constitute the `environment') in terms
of the lower-order ones by functional relations (`slaving' or `factorization')
induces dissipation in the dynamics of the subsystem driven by the stochastic
fluctuations of the environment, which we call the `correlation  noises'.
These two aspects are related by the
fluctuation-dissipation relation. This is the quantum field equivalent of
the BBGKY hierarchy in Boltzmann's theory. Any subsystem involving a finite
number of correlation functions defines an effective theory,
which is, by this reasoning, intrinsically dissipative.
The relation of loop expansion and correlation
order is expounded. We see that ordinary quantum field theory which involves
only the mean field and a two-point function, or any finite-loop effective
action in a perturbative theory are, by nature, effective theories
which possess these properties.
Histories defined by lower-order correlation functions can be decohered by the
noises from the higher order functions and acquire classical stochastic
attributes.  The present scheme invoking the correlation order is a
natural way to describe the quantum to classical transition for a closed
system as it avoids {\it ad hoc} stipulation of the system-environment split.
It is through decoherence that the subsystem variables become classical
and the subdynamics becomes stochastic.

\subsubsection{Key Points}

{\it Fluctuations in Mean Field and Correlations;
Lessons from Critical Dynamics and Kinetic Theory}\\

One useful departure point from the conventional treatment is to view
the so-called `mean' field not as the actual expectation value of the field,
but rather as representing the local value of the field within one particular
history. Quantum evolution encompasses the coherent superposition
of all possible histories \cite{conhis}.
These quantities are subject to fluctuations.
The theory may be enlarged by including some
correlation functions as independent variables along with the `mean'
field. These correlation functions will be subject to fluctuations of
their own. We can recognize the inadequacy of theories built on mean
field and its fluctuations from examing some familiar physics problems.

One area where the correlations of fields play a central role is
critical dynamics. There, 
one focuses on the evolution of some `order parameter' whose mean value obeys
a Landau - Ginzburg equation \cite{Landau}. In practice, however,
the actual value of the order parameter can have strong fluctuations around
the mean value as the critical point is approached, and the Landau - Ginzburg
equation must be generalized to include stochastic terms such as is used in
the Cahn - Hilliard theory \cite{cahnhilliard}. The difference in our
treatment is that the stochastic source (noise) term is not put in by hand,
but is derived from the  given structures of the field theory in question.
Equivalently, it is possible to describe the evolution in
terms of a Fokker - Plank equation \cite{langer1}, which may be used
to derive a hierarchy of equations for the correlation functions.
For practical applications this hierarchy must be truncated, as was done
in Langer's treatment \cite{langer2}. One of us has used the CTP nPI
method in quantum field theory to the problem of spinodal decomposition
before \cite{sdqft}. Although the effect of fluctuations on the correlation
functions has not been explored
in full (see, however, \cite{GraGre,OCSrg}),
this step is nevertheless common in dilute gas dynamics \cite{KacLogan}.

In the dynamics of a dilute gas \cite{Akhiezer,Balescu} the exact
Newton's or Hamilton's equations for the
evolution of a many body system may be translated into a Liouville
equation for the distribution function or the BBGKY hierarchy for the
sequence of partial (n- particle) distributions. This reformulation is
only formal, which involves no loss of information or predictability.
A more realistic description of the dynamics corresponding to physical
conditions comes from a truncated BBGKY hierarchy, where the higher order
distributions are substituted by functionals of the lower order, and
ultimately, the one- particle distributions.
Constructed perturbatively, this effective theory follows
only approximately the actual dynamics.
Moreover, these functionals embody some relevant boundary conditions (such as
the `weakening of correlations' hypothesis \cite{Akhiezer}), which make
them noninvariant upon time reversal. This is how dissipation in the explicitly
irreversible Boltzmann's equation appears.
On closer examination, it is seen that
the one- particle distribution function itself describes only the mean
number of particles within a certain location in phase space; the actual
number is also subject to fluctuations. From the average size of the
equilibrium
fluctuations, which can be determined from Einstein's formula, and the
dissipative element of the dynamics, which is contained in the collision
integral, it is possible to compute the stochastic driving force consistent
with the fluctuation- dissipation relation near equilibrium \cite{KacLogan}.\\


{\it Dyson-Schwinger Hierarchy; the Master Effective Action,
Truncation and Slaving}\\

We want to describe a quantum field in terms of the mean field and the
(infinite number of) correlation functions.
Our starting point shall be the well-known fact
that the set of all Wightman functions (time ordered products of field
operators) determines completely the quantum state of a field \cite{Haag}.
Instead of following the evolution of the field in any of the conventional
representations (Schr\"odinger, Heisenberg or Dirac's), we shall
focus on the dynamics of the full hierarchy of Wightman functions. To this
end it is convenient to adopt Schwinger's ``closed time-path'' techniques
\cite{ctp}, and consider time ordered Green functions as a subset of all
Green functions path- ordered along a closed time loop (see below). The
dynamics of this larger set is described by the Dyson - Schwinger equations
\cite{ItzykZuber}.

We will first show that the Dyson- Schwinger hierarchy may be  obtained via the
variational principle from a functional which we call the `Master Effective
Action' (MEA). This is a formal action functional where each Wightman function
enters as an independent variable.
We will then show that any field theory based on a finite number of (mean
field plus ) correlation functions can be viewed as a subdynamics of the
Dyson- Schwinger hierarchy. The specification
of a subdynamics involves two steps: first, a finite set of variables from
the original hierarchy is identified to be the `relevant' \cite{projop}
variables (which constitute the subsystem). Second, the remaining
`irrevelant' (or `environment') variables are {\it slaved} to the
former. Slaving (or `factorization' in the Boltzmann theory)
means that irrelevant variables are substituted by set
functionals of the relevant variables.  The process of extraction of a
subdynamics from the Dyson- Schwinger hierarchy has a correlate at the level
of the effective action,
where the MEA is truncated to a functional of a finite number of variables.
The finite effective actions so obtained (influence action \cite{if})
are generally nonlocal and complex, which is what gives rise to the noise
and dissipation in the subdynamics.
Moreover, since the slaving process generally involves the choice of an
arrow of time, it leads to irreversibility in the cloak of dissipation in the
subdynamics \cite{projop}.\\

{\it Decoherence and Noise, Fluctuations and Dissipation}\\

Under realistic conditions, one may not be as much
concerned with the full quantum evolution of the field as with the
development of `classical' theories where fields are described as c-numbers,
plus perhaps a small number of correlation functions to keep track of
fluctuations. These classical theories represent the physically
observable dynamics after the process of decoherence \cite{envdec,conhis}
has destroyed or diminished the coherence of the field. In any case, no
actual observation could disclose the infinite number of degrees of freedom
of the quantum field, and therefore any conceivable observational situation
may be described in the language of a suitable complex `classical'
theory in this sense.

Decoherence is brought by the effect of a coarse-grained environment
(or `irrevelant' sector) on the system (or the `relevant' sector).
In simple models, this split is imposed by hand, as when some of the fields,
or the field values within a certain region of spacetime, are chosen as
relevant. Here, we will follow the approach of our earlier work on
the decoherence of correlation histories \cite{DCH}. There is no need
to select {\sl a priori} a relevant sector within the theory.
Instead, we shall seek a natural criterium for successive truncations in
the hierarchy of correlations, the degree of truncation depending on the
stipulated accuracy of measurement which can be carried out on the system.
In this framework, decoherence occurs as a consequence of the fluctuations
in the higher order correlations and results in a classical dissipative
dynamics of the lower order correlations.

Here, we adopt the consistent histories formulation of quantum mechanics
\cite{conhis} for the study of the quantum to classical transition problem.
We consider the full evolution of the field described by the
Dyson- Schwinger hierarchy as a fine- grained history while histories
where only a
finite number of Wightman functions are freely specified (with all others
slaved to them) are therefore coarse- grained. We have shown that the
finite effective actions obtained for the subsystems of lower-order
correlations are related to the decoherence functional between two such
histories of correlations \cite{DCH}, its acquiring an imaginary part
signifies the existence of noise which facilitates decoherence.
Thus decoherence of correlation histories is a necessary condition for the
relevance of the c-number theory as a description of observable phenomena.\\
It can be seen that if the c-number theory which emerges from the quantum
subdynamics  is dissipative, then it must also be stochastic.
\footnote{ Because the fundamental variables are quantum in nature,
and therefore subject to fluctuations, a classical, dissipative dynamics
would demand the accompaniment of stochastic sources (in agreement with the
`fluctuation - dissipation theorems', for, otherwise, the theory would
permit unphysical phenomena as the damping away of zero - point fluctuations.)
Of course, these uncontrollable fluctuations may be seen at
the origin of many phenomena where structure seems to spring `out of nothing',
such as the nucleation of inhomogeneous true vacuum bubbles in a supercooled
false vacuum, or the development of inhomogeneities out of a homogeneous early
universe.} From our correlation history viewpoint, the stochasticity is in fact
not confined to the field distributions-- the correlation functions would
become
stochastic as well \cite{KacLogan}.

Following Feynman and others \cite{if} in their illustration of
how noise can be defined from the influence functional, we can
relate the imaginary part of the finite effective actions
describing the truncated correlations to the auto-correlation
of the stochastic sources, i.e., correlation noises, which  drive the
c-number fields and their correlation functions via the  Langevin- type
equations. From the properties of the complete (unitary) field theory
which constitutes the  closed (untruncated) system, one can show that
the imaginary part of the effective action is related  to the nonlocal
part of the real part of the effective action which depicts dissipation.
This is of course where the fluctuation- dissipation theorem for
non-equilibrium systems originates \cite{fdrsc}.

We thus see once again the intimate connection amongst the three aspects
of the theory, decoherence, dissipation, and fluctuations
\cite{HuTsukuba,GelHar2}, now manifesting in the hierarchy of correlations
which defines the subsystems.\\

This paper is organized as follows.  The next subsection
contains a  brief statement of the fluctuation- dissipation
theorem in two variant forms, described here to clarify our meaning of this
relation.
Section 2 is a  quick overview of functional  methods  in
causal  quantum field  theory  based on \cite{ch87,ch88}
(See references therein for sources and details.)

In  Section 3 we present a formal  construction  of  the
Master  Effective  Action. We compute it with the background  field method,
and discuss its relation with the finite theories by truncation and slaving.
In Section 4 we introduce decoherence  in  histories and
discuss the relationship of the master effective  action  and its
truncations to the decoherence functional.
As illustration, we give a brief description of the
decoherence of the mean field in a self interacting field theory.  This
example is along the line of development in \cite{nfsg} and \cite{fgic}.

Section  5 is  devoted  to  the new concept  of
`correlation  noise',  which measures the  fluctuations  of the
propagators  themselves  due  to the truncation of higher correlations.
We present a simple example where `correlation noise' can be  derived
through  direct  physical  arguments,  and  then show how these results
follow rigorously as  a simple case of the formal theory developed
in sections 3 and 4.

We conclude the paper with some discussion of possible applications of this
formalism and viewpoint in effective field theory, and quantum statistical
processes in the early universe and black holes in Section 6.

\subsection{The Fluctuation - Dissipation Relation (FDR)}

Given the relevance of the FDR to the discussion below, it is useful to
spend a few moments clarifying our meaning. Let us consider a near
equilibrium system described by a single variable $x$; let the entropy of
the system be $S(x)$, and call $y=-dS/dx$ the corresponding `thermodynamic
force'. Let $x$ vanish in equilibrium. A macroscopic fluctuation can be
absorbed into a redefinition of the equilibrium;
as a first order approximation, we may write
$\dot x=-ay$. The condition that entropy should be non-decreasing in the
process implies that $a$ should be positive. A microscopic fluctuation which
induces deviations from the average constitutes a stochastic force $j$,
yielding a stochastic equation $\dot x=-ay+j$.
One can close the system of evolution equations by assuming
the linear law $y=bx$. Again, because $S$ is maximum at the origin, $b\ge 0$.
We thus have a Langevin equation
$\dot x=-cx+j$, where $c=ab$ is positive. This leads immediately to

\begin{equation}
\label{FDT1}x(t)=x(0)e^{-ct}+\int^t_0dt^{\prime}~e^{-c(t-t^{\prime})}j(t^{%
\prime})
\end{equation}

In the literature there are two different results under the heading of
`Fluctuation - Dissipation Theorem', obtained by manipulating Eq. (\ref
{FDT1}) in two different ways. The `Landau - Lifschitz' result \cite{Landau}
is obtained by multiplying this equation by $x(0)$, taking the average over
the stochastic forces $j$ (assuming the average $\langle x(0)j(t)\rangle =0$),
and integrating over time, leading to the Green - Kubo formula

\begin{equation}
\label{FDT2}c^{-1}=(\langle x^2(0)\rangle )^{-1}\int_0^{\infty}dt~\langle
x(0)x(t)\rangle
\end{equation}

In this work, we shall be more concerned with the second customary
formulation of the FDR, which goes back to Callen and Welton's original work
\cite{fdt}. Squaring both sides of Eq. (\ref{FDT1}), and taking the
ensemble average, assuming as before that $\langle x(0)j(t)\rangle =0$,
$\langle j(t)j(t^{\prime })\rangle =\sigma ^2\delta (t-t^{\prime}) $ and
that $\langle x^2(t)\rangle $ remains independent of $t$
at its equilibrium value, we find

\begin{equation}
\label{FDT3}\sigma^2=2c\langle x^2\rangle
\end{equation}
Thus, the FDR determines the statistics of the random source $j$ from the
given dissipation coefficient $c$ and the equilibrium fluctuations.

The actual result quoted by Callen and Welton reads as follows. Define at
each frequency the impedance $Z$ from $j(\omega )=i\omega Zx(\omega )$, and
the resistance $R={\rm Re}Z$. In general the noise $j(t)$ will be colored,
but time translation invariant, in the sense that $<j(\omega )j(\omega
^{\prime })>\sim \delta (\omega +\omega ^{\prime })$. Let us call $\sigma
^2\equiv <j(t)^2>$. Then Callen and Welton's FDR states that

\begin{equation}
\label{FDT4}\sigma ^2= \frac 2 \pi  \int_0^\infty d\omega
\;\;R\left( \omega \right) f\left( \omega ,T\right)
\end{equation}
where $f$ is an universal function depending only on temperature

\begin{equation}
\label{FDT5}f\left( \omega ,T\right) =\hbar \omega \left( \frac 12+
\frac 1{ e^{\frac{\hbar \omega }{kT}}-1 }\right)
\end{equation}

In quantum field theory this formula is equivalent to the KMS condition
\cite{kms}. Indeed, from the definition of the impedance we have

\be
\sigma ^2={\frac 1{(2\pi )}}\int_0^\infty d\omega \omega ^2\left| Z(\omega
)\right| ^2G_1(\omega );
\te
the KMS condition further links the Hadamard and the Jordan propagators,
namely

\be
G_1={\frac{2}{\omega}}f(\omega ,T)G(\omega ){\rm sgn}(\omega )
\te
Finally, from $G_{ret}=-iG\theta (t-t^{\prime })$ and $G_{ret}=(i\omega
Z)^{-1}$ we conclude

\be
G(\omega )={\frac{2{\rm sgn}(\omega )}\omega }
\frac R{\left| Z(\omega)\right| ^2}),
\te
just as dictated by the Callen and Welton expression.

\section{Functional Methods and the Dyson - Schwinger Hierarchy}

Our goal in this section is to show how a formal Effective Action may be
defined, which generates the whole hierarchy of Dyson - Schwinger's equations%
\cite{ItzykZuber}. We shall use Schwinger's CTP techniques \cite{ctp},
extending earlier works on higher-particle-irreducible effective actions
\cite{2pi}. We shall begin by recapitulating the train of thought which led
us to this particular formulation of quantum field theory.

The effective action usually appears in quantum field theory textbooks in the
context of perturbative expansions \cite{ramond}. As it is by now a
common procedure, an efficient way to generate the Feynman Green
functions of a quantum field is from the variational derivatives of a
generating functional. The generating functional itself admits the path
integral representation

\begin{equation}
\label{R1}e^{iW[J]}=\int~D\Phi~e^{i \{S[\Phi]+\int d^4xJ(x)\Phi (x)\}}
\end{equation}
The coefficients in the Taylor expansion of $W$ with respect to $J$ are the
connected Feynman graphs of the (scalar) field theory $\Phi$. In particular,
we may introduce the `background' or `mean' field

\begin{equation}
\label{R2}\phi (x)={\frac{\delta W[J]}{\delta J}}
\end{equation}
A more efficient representation of the Green functions is afforded
by the Legendre transformation of $W$

\begin{equation}
\label{R3}\Gamma [\phi ]=W[J]-\int d^4x~J(x)\phi (x)
\end{equation}
since the Taylor expansion of $\Gamma $ involves only one particle
irreducible (1PI) Feynman graphs. Actually, the physical meaning of the
`effective action' $\Gamma $ goes beyond this: for a constant background,

\begin{equation}
\label{R4}\Gamma [\phi ]=-\int ~d^4x~V[\phi ],
\end{equation}
and the `effective potential' $V$ is the actual free energy density of the
quantum field. Thus it is tempting to consider the inverse of (\ref{R2})
as the equation of motion for the background field

\begin{equation}
\label{R5}{\frac{\delta\Gamma}{\delta\phi (x)}}=-J(x)
\end{equation}
However, this equation is generally complex. This property follows from the
fact that the `background' field is not a true expectation value, but
rather a matrix element of the field between IN and OUT states
\cite{hartlehu}.

We can get around this difficulty by adopting Schwinger's `closed
time-path' formalism. Let us couple the field to an external source $J(x)$,
and adopt the interaction picture where states evolve according to the
evolution operator

\begin{equation}
\label{R5.1}U(t)=T(e^{i\int_{-\infty}^tJ(x)\Phi (x)})
\end{equation}
where $T$ stands for time ordering. Then, if the state in the distant past is
$\vert IN\rangle$ (not necessarily a vacuum state), the expectation value of
the field at time $t$ is

\begin{equation}
\label{R6}\phi (x)=\langle IN|\tilde T(e^{-i\int^td^4x^{\prime
}~J(x^{\prime })\Phi (x^{\prime })})\Phi (x)T(e^{i\int d^4x^{\prime \prime
}~J(x^{\prime \prime })\Phi (x^{\prime \prime })})|IN\rangle |
\end{equation}
where $\tilde T$ stands for anti-temporal ordering. This may be rewritten as

\begin{equation}
\label{R6.2}\phi (x)={\frac{\delta}{\delta J(x)}}e^{iW[J,J']}%
\vert_{J'\equiv J}
\end{equation}
where $W$ is  the CTP generating functional

\begin{equation}
\label{R6.1}e^{iW[J,J']}=\langle IN|\tilde T(e^{-i\int d^4x^{\prime
}~J'(x^{\prime })\Phi (x^{\prime })})T(e^{i\int d^4x^{\prime \prime
}~J(x^{\prime \prime })\Phi (x^{\prime \prime })})|IN\rangle
\end{equation}
The time integrals now extend to the far future. The generating
functional admits a path integral representation,

\begin{equation}
\label{R8}e^{iW[J,J']}=
=\int~D\Phi~D\Phi'~e^{i\{S[\Phi]-S[\Phi']+\int d^4x~(J(x)\Phi (x)
-J'(x)\Phi' (x)\}}
\end{equation}
Thus, we have been led to formally doubling the degrees of freedom and
integrating over pairs of histories $(\Phi,\Phi')$ which converge in the
far future. In condensed notation

\begin{equation}
\label{R9}e^{iW[J_a]}=\int~D\Phi^a~e^{i(S[\Phi^a]+J\Phi )}
\end{equation}
where the index $a$ indicates a doubling into umprimed and primed
quantities (with an opposite sign for the latter).
$S[\Phi^a]=S[\Phi]-S[\Phi']^{*}$ (complex conjugation
applies if an $i\epsilon $ term has been included to enforce the boundary
conditions), and the `internal' index $a$ is lowered with the `metric'
$g_{ab}={\rm diag}(1,-1)$. We have  also omitted summation signs over
continuous
indexes, thus

\begin{equation}
\label{R10}J\Phi\equiv\int~d^4x~J_a(x)\Phi^a(x)
\end{equation}

We can actually define two background fields by taking variations with
respect to either source; the physical situation corresponds to the case
when both sources agree.
The closed time-path effective action (CTPEA) is the double Legendre
transform of the generating functional

\begin{equation}
\label{R11}\Gamma [\phi^a]=W[J_a]-\int d^4x~J_a(x)\phi^a (x)
\end{equation}
The equations of motion now form a coupled system

\begin{equation}
\label{R12}{\frac{\delta\Gamma}{\delta\phi^a (x)}}=-J_a(x)
\end{equation}
They are identical to each other when $J=J'$ is the
physical external source, and they admit a solution where $\phi=\phi'$
is the physical mean field. Of course, both background fields must be
identified only after the variational derivative has been computed \cite{ch87}.
As opposed to the IN-OUT formulation, the equations of motion (\ref{R12})
are always real and causal whereas in cases where the IN-OUT effective
action becomes complex, the IN-IN effective action leads to dissipative
evolution. This fact underlines that the doubling of degrees of freedom was,
after all, essential, since no action functional of the physical mean field
alone could possibly lead to the right (dissipative) equations of motion.

Breaking of time symmetry in the mean field equations of motion also points
to the fact that the mean field cannot be considered as a closed system.
Indeed, as we get to understand the problem better, and as we shall show in
this
and related work, ordinary quantum field theory based on mean field and
fluctuations is an open system. This is because the mean field interacts,
in any non-linear field theory, with its own quantum (and thermal)
fluctuations, which act as an `environment' \cite{ch89}.
In principle it is possible to shift the division between the system
and its environment by including
into the former some of the correlations describing those fluctuations. This
is analogous to Langer's approach to critical dynamics \cite{langer2}, and
has been carried out in field theory by many authors, namely Dahmen and Jona
- Lasinio, Cornwall, Jackiw and Tomboulis, Norton \cite{2pi}, and, in the
context of the CTP approach, the present authors  \cite{ch88}  \cite{sdqft}.
Not surprisingly, this step leads to a further compression of the
perturbative series; by including two point functions as independent
variables, we obtain a generating functional whose Taylor expansion contains
only two-particle irreducible (2PI) Feynman graphs.
Generalization of this formalism shall be described in detail in the next
section.

However, there is a physical issue which is not resolved by extending
to higher irreducible effective actions, namely, how to describe the dynamics
of the field from the vantage point of a local observer.
A possible answer is suggested by physical reasoning inspired by Onsager and
Landau, namely, we obtain the relevant dynamics by including stochastic
terms in the mean field equations of motion. These stochastic sources
generate dissipative dynamics in the mean field equations which are related
to the known quantum and thermal fluctuations of the field
through the fluctuation- dissipation relations.
The main thrust of this paper is precisely to show how the CTP
effective action formalism may be used to explicate these physical ideas and
to construct a consistent theory valid beyond the linear response
regime (no equilibrium initial conditions need be assumed).

Recognition of the stochastic nature of semiclassical evolution also
helps to dispel the appearance of redundancy in the CTPEA. Let us observe
that quite generally the CTPEA obeys $\Gamma [\phi ,\phi ]=0$ and $\Gamma
[\phi,\phi']=-\Gamma [\phi,\phi']^{*}$. Therefore, if we separate
the CTPEA in its real and imaginary parts, we find that the former is an odd
function of $\phi$ and $\phi'$, while the latter is even. If we
consider only linearized fluctuations $\Delta \phi ^a$ around an extremum of
the CTPEA, i. e., a solution of the mean field equations of motion, the
CTPEA must take the form

\begin{equation}
\label{R14}\Gamma [\Delta\phi^a] ={\frac{1}{2}}\int~d^dxd^dx^{\prime}\{-[%
\Delta\phi ](x){\cal D}(x,x^{\prime})
\{\Delta\phi\}(x^{\prime})+i[\Delta\phi](x)N(x,x^{\prime}) [\Delta\phi
](x^{\prime})\}
\end{equation}
where $[\Delta\phi ]=(\Delta\phi-\Delta\phi')$, $\{\Delta\phi\}=(\Delta%
\phi+\Delta\phi')$, and ${\cal D}$ and $N$ are two non-local kernels, to
be further discussed below. The seeming redundancy appears because $N$ does
not contribute to the mean field equations of motion, which take the simple
form

\begin{equation}
\label{R15}\int~d^dx^{\prime}{\cal D}(x,x^{\prime})\Delta\phi(x^{\prime})=0
\end{equation}
after identifying $\Delta \phi'=\Delta \phi $. The answer
lies, of course, in that $N$ contains the information on the {\sl departure}
of the actual evolution from mean field behavior; this deviation may be
accounted for by coupling the fluctuations to a stochastic external force
 \cite{if}. We shall return to this point after introducing the
Decoherence Functional below.

Bringing together all the different strands, we arrive at the conclusion
that a suitable formulation of quantum field theoretic evolution should
incorporate both the CTP technique, to allow for causal evolution, and the
higher particle irreducible techniques, to obtain a self consistent dynamics
of correlations. We will see that any quantum field theory based on a
finite subset of correlation functions (which defines an open system --
the truncation rendering it an effective theory) is  necessarily dissipative.
To obtain a complete and exact description of the full dynamics (of the
closed system) we must embrace {\sl all} correlation functions.
One can actually construct a `master' effective action for this purpose,
as we now proceed to show.

\section{The Master Effective Action}

Our goal in this section is to show that it is (formally) possible to write
down a functional of a c-number background field and a string of Green
functions, such that variation of this functional yields the usual Dyson -
Schwinger equations of QFT. For simplicity, we shall study a scalar field
theory; since we are interested in the CTP formulation of the theory,
however, the field must be thought of as carrying an internal index $a$,
denoting the umprimed and primed quantities respectively,
which is raised or lowered with a `metric' $g_{ab}={\rm diag}(1,-1)$.
Also, CTP boundary conditions must be understood in all
equations.

\subsection{Formal Construction of the Master Effective Action}

We consider then a scalar field theory whose action

\begin{equation}
S[\Phi ]={\frac{1}{2}}S_2\Phi^2 +S_{int}[\Phi ]
\end{equation}
decomposes into a free part and an interaction part

\begin{equation}
S_{int}[\Phi ]=\sum_{n=3}^{\infty}{\frac{1}{n!}}S_n\Phi^n
\end{equation}
Here and after, we use the shorthand

\begin{equation}
K_n\Phi^n\equiv\int~d^dx_1...d^dx_n~K_{na^1...a^n}(x_1,...x_n)
\Phi^{a^1}(x_1)...\Phi^{a^n}(x_n)
\end{equation}
where the kernel $K$ is assumed to be totally symmetric.

Let us define also the `source action'

\begin{equation}
J[\Phi ]=J_1\Phi +{\frac{1}{2}}J_2\Phi^2+J_{int}[\Phi ]
\end{equation}
where $J_{int}[\Phi ]$ contains the higher order sources

\begin{equation}
J_{int}[\Phi ]=\sum_{n=3}^{\infty}{\frac{1}{n!}}J_n\Phi^n
\end{equation}
and define the generating functional

\begin{equation}
Z[\{J_n\}]=e^{iW[\{J_n\}]}=\int~D\Phi~e^{iS_t[\Phi ,\{J_n\}]}
\end{equation}
where

\begin{equation}
S_t[\Phi ,\{J_n\}]=J_1\Phi +{\frac{1}{2}}(S_2+J_2)\Phi^2+S_{int}[\Phi ]
+J_{int}[\Phi ]
\end{equation}
We shall also call

\begin{equation}
S_{int}[\Phi ]+J_{int}[\Phi ]=S_I
\end{equation}

As it is well known, the Taylor expansion of $Z$ with respect to $J_1$
generates the expectation values of path - ordered products of fields

\begin{equation}
{\frac{\delta^n Z}{\delta J_{1a^1}(x_1)...\delta J_{1a^n}(x_n)}}= \langle
P\{\Phi^{a^1}(x_1)...\Phi^{a^n}(x_n)\}\rangle\equiv F_n^{a^1...a^n}
(x_1,...x_n)
\end{equation}
while the Taylor expansion of $W$ generates the `connected' Green
functions (`linked cluster theorem' \cite{Haag})

\begin{equation}
{\frac{\delta^n W}{\delta J_{1a^1}(x_1)...\delta J_{1a^n}(x_n)}}= \langle
P\{\Phi^{a^1}(x_1)...\Phi^{a^n}(x_n)\}\rangle_c\equiv C_n^{a^1...a^n}
(x_1,...x_n)
\end{equation}
Comparing these last two equations, we find the rule connecting the $F$'s
with the $C$'s. First, we must decompose the ordered index set $(i_1,...i_n)$
($i_k=(x_k,a^k)$) into all possible clusters $P_n$. A cluster is a partition
of $(i_1,...i_n)$ into $N_{P_n}$ ordered subsets $p=(j_1,...j_r)$. Then

\begin{equation}
F_n^{i_1...i_n}=\sum_{P_n}\prod_{p}C_r^{j_1...j_r}
\end{equation}
Now from the obvious identity

\begin{equation}
{\frac{\delta Z}{\delta J_{ni_1...i_n}}}\equiv \frac{1}{n!}
{\frac{\delta^n Z}{\delta J_{i_1}...\delta J_{i_n}}}
\end{equation}
we obtain the chain of equations

\begin{equation}
{\frac{\delta W}{\delta J_{ni_1...i_n}}}\equiv {\frac{1}{n!}}
\sum_{P_n}\prod_{p}C_r^{j_1...j_r}
\end{equation}

We can invert these equations to express the sources as functionals of the
connected Green functions, and define the master effective action (MEA)
as the full Legendre transform of the connected generating functional

\begin{equation}
\Gamma_{\infty}[\{C_r\}]=W[\{J_n\}]-\sum_n{\frac{1}{n!}}J_n\sum_{P_n}%
\prod_pC_r
\end{equation}
The physical case corresponds to the absence of external sources, whereby

\begin{equation}
{\frac{\delta\Gamma_{\infty}[\{C_r\}]}{\delta C_s}}=0
\end{equation}
This hierarchy of equations is equivalent to the Dyson- Schwinger series.

\subsection{The Master Effective Action and the Background Field Method}

The master effective action just introduced becomes more manageable if one
applies the background field method (BFM) \cite{effaction} approach. We first
distinguish the mean field and the two point functions

\begin{equation}
C_1^i\equiv\phi^i
\end{equation}

\begin{equation}
C_2^{ij}\equiv G^{ij}
\end{equation}
We then perform the Legendre transform in two steps: first with respect to
$\phi$ and $G$ only, and then with respect to the rest of the Green functions.
The first (partial) Legendre transform yields

\begin{equation}
\Gamma_{\infty}[\phi, G,\{C_r\}]\equiv\Gamma_2[\phi, G,\{j_n\}]- \sum_{n\ge
3}{\frac{1}{n!}}J_n\sum_{P_n}\prod_pC_r
\end{equation}
Here $\Gamma _2$ is the two particle-irreducible (2PI) effective action
\cite{2pi}

\begin{equation}
\Gamma_2[\phi, G,\{J_n\}]=S[\phi]+{\frac{1}{2}}G^{jk}S,_{jk} -{\frac{i}{2}}
\ln ~{\rm Det}~G+J_{int}[\phi ] +{\frac{1}{2}}G^{jk}J_{int,jk}+W_2
\end{equation}
and $W_2$ is the sum of all 2PI vacuum bubbles of a theory whose action is

\begin{equation}
S^{\prime}[\varphi ]={\frac{i}{2}}G^{-1}\varphi^2+S_Q[\varphi ]
\end{equation}

\begin{equation}
S_Q[\varphi ]=S_I[\phi +\varphi ]-S_I[\phi ]-S_I[\phi ],_i\varphi^i -
{\frac{1}{2}}S_I[\Phi ],_{ij}\varphi^i\varphi^j
\end{equation}
where $\varphi$ is the fluctuation field around $\phi$, i.e., $\Phi = \phi +
\varphi$.
Decomposing $S_Q$ into source-free and source-dependent parts, and Taylor
expanding with respect to $\varphi $, we may define the background-field
dependent coupling and sources

\begin{equation}
S_Q[\varphi ]=\sum_{n\ge 3}{\frac{1}{n!}}(\sigma_n+\chi_n)\varphi^n
\end{equation}
where

\begin{equation}
\sigma_{ni_1...i_n}=\sum_{m\ge n}{\frac{1}{(m-n)!}}
S_{mi_1...i_nj_{n+1}...j_m}\phi^{j_{n+1}}...\phi^{j_m}
\end{equation}

\begin{equation}
\chi_{ni_1...i_n}=\sum_{m\ge n}{\frac{1}{(m-n)!}}
J_{mi_1...i_nj_{n+1}...j_m}\phi^{j_{n+1}}...\phi^{j_m}
\end{equation}
Now, from the properties of the Legendre transformation, we have, for $n>2$,

\begin{equation}
{\frac{\delta W}{\delta J_{n}}}\vert_{J_1,J_2}\equiv
{\frac{\delta\Gamma_{\infty}}{\delta J_{n}}}\vert_{\phi ,G}
\end{equation}
Computing this second derivative explicitly, we conclude that

\begin{equation}
{\frac{\delta W}{\delta J_{n}}}\vert_{J_1,J_2}\equiv {\frac{1}{n!}}\phi^n
+{\frac{1}{2~(n-2)!}}G\phi^{n-2}+\sum_{m=3}^n {\frac{\delta\chi_m}{\delta J_n}}
{\frac{\delta W_2}{\delta\chi_{m}}}
\end{equation}
Comparing this equation with

\begin{equation}
{\frac{\delta W}{\delta J_{ni_1...i_n}}}\equiv {\frac{1}{n!}}
\sum_{P_n}\prod_{p}C_r^{j_1...j_r}
\end{equation}
we obtain the identity

\begin{equation}
{\frac{\delta W_2}{\delta\chi_{ni_1...i_n}}}\equiv {\frac{1}{n!}}
\sum^*_{P_n}\prod_{p}C_r^{j_1...j_r}
\end{equation}
where the * above the sum means that clusters containing one element subsets
are deleted. This and the equality

\begin{equation}
\sum_{n\ge 3}{\frac{1}{n!}}J_n\sum_{P_n}\prod_pC_r=J_{int}[\phi ]+
{\frac{1}{2}}
G^{ij}{\frac{\delta J_{int}[\phi ]}{\delta\phi^i\delta \phi^j}}
+\sum_{n\ge 3}{\frac{1}{n!}}\chi_n\sum ^*_{P_n}\prod_pC_r
\end{equation}
allow us to write
\bea
\Gamma _\infty [\phi ,G,\{C_r\}]  &\equiv & \ S[\phi ]+(
{\frac 12})G^{ij}{\frac{\delta S[\phi ]}{\delta \phi ^i\delta \phi ^j}}
-{\frac i2}\ln ~{\rm Det}~G  \nn \\
& & +\{W_2[\phi ,\{\chi _n\}]-\sum_{n\ge 3}{\frac 1{n!}}\chi_n
\sum_{P_n}^{*}\prod_pC_r\}
\tea

This entails an enormous simplification, since it implies that to compute $%
\Gamma_{\infty}$ it is enough to consider $W_2$ as a functional of the $%
\chi_n$, without ever having to decompose these background dependent sources
in terms of the original external sources.

\subsection{Truncating the Master Effective Action: Loop Expansion and
Correlation Order}

After obtaining the formal expression for $\Gamma _\infty $, and thereby the
formal hierarchy of Dyson - Schwinger equations, we should proceed with it
much as with the BBGKY hierarchy in statistical mechanics \cite{Huang},
namely, truncate it and close the lower-order equations
by constraining the high order correlation functions to be given
(time-oriented) functionals of the lower correlations. Truncation proceeds
by discarding the higher correlation functions and replacing them by given
functionals of the lower ones, which represent the dynamics in some
approximate sense \cite{Akhiezer}. The system which results is an open system
and the dynamics becomes an effective dynamics.

It follows from the above that truncations will be generally related to
approximation schemes. In field theory we have several such schemes
available, such as the loop expansion, large $N$ expansions, expansions in
coupling constants, etc. For definiteness, we shall study the case of the
loop expansion, although similar considerations will apply to any of the
other schemes.

Taking then the concrete example of the loop expansion, we observe that the
nonlocal $\chi$ sources enter into $W_2$ in  as  many nonlinear couplings of
the fluctuation field $\varphi$. Now, $W_2$ is given by a sum of connected
vacuum bubbles, and any such graph satisfies the constraints

\begin{equation}
\sum nV_n=2i
\end{equation}

\begin{equation}
i-\sum V_n=l-1
\end{equation}
where $i,l,V_n$ are the number of internal lines, loops, and vertices with $%
n $ lines, respectively. Therefore,

\begin{equation}
l=1+\sum{\frac{n-2}{2}}V_n
\end{equation}
we conclude that $\chi_n$ only enters the loop expansion of $W_2$ at order $%
n/2$. At any given order $l$, we are effectively setting $\chi_n\equiv 0$, $%
n>2l$. Since $W_2$ is a function of only $\chi_3$ to $\chi_{2l}$,
it follows that the $C_r$'s cannot be all independent. Indeed, the equations
relating sources to Green functions

\begin{equation}
{\frac{\delta W_2}{\delta\chi_{ni_1...i_n}}}\equiv {\frac{1}{n!}}
\sum^*_{P_n}\prod_{p}C_r^{j_1...j_r}
\end{equation}
have now turned, for $n>2l$, into the algebraic constraints

\begin{equation}
\sum ^*_{P_n}\prod_{p}C_r^{j_1...j_r}\equiv 0
\end{equation}
In other words, the constraints which makes it possible to invert the
transformation from sources to Green functions allow us to write the higher
Green functions in terms of lower ones. In this way, we see that the loop
expansion is by itself a truncation in the sense above and hence any
finite loop or perturbation theory is intrinsically an effective theory.

Actually, the number of independent Green functions at a given number of
loops is even smaller than $2l$. It follows from the above that $W_2$ must
be linear on $\chi_n$ for $l+2\le n\le 2l$. Therefore the corresponding
derivatives of $W_2$ are given functionals of the $\chi_m$, $m\le l+1$.
Writing the lower sources in terms of the lower order Green functions, again we
find a set of constraints on the Green functions, rather than new equations
defining the relationship of sources to functions. These new constraints
take the form

\begin{equation}
\sum ^*_{P_n}\prod_{p}C_r^{j_1...j_r}=f_n(G,C_3,...C_{l+1})
\end{equation}
for $l+2\le n\le 2l$. In other words, to a given order $l$ in the loop
expansion, only $\phi$, $G$ and $C_r$, $3\le r\le l+1$, enter into
$\Gamma_{\infty}$ as independent variables. Higher correlations are expressed
as functionals of these by virtue of the constraints implied by the loop
expansion on the functional dependence of $W_2$ on the sources.

However, these constraints are purely algebraic, and therefore do not define
an arrow of time. The dynamics of this lower order functions is unitary.
Irreversibility appears only when one makes a time-oriented ansatz in the
form of the higher correlations,
such as the `weakening of correlations' principle invoked in the
truncation of the BBGKY hierarchy \cite{Akhiezer}. This is done by
substituting some of the allowed correlation functions at a given number of
loops $l$, by solutions of the $l$-loop equations of motion. Observe
that even if we use exact solutions, the end result is an irreversible
theory, because the equations themselves are only an approximation to the
true Dyson - Schwinger hierarchy.

To summarize, the truncation of the MEA in a loop expansion scheme proceeds
in two stages. First, for a given accuracy $l$, an $l$- loop effective
action is obtained which depends only on the lowest $l+1$ correlation
functions, say, $\{\phi ,G,C_3,\ldots C_{l+1}\}$.
This truncated effective action generates the $l$-loop equations
of motion for these correlation functions. In the second stage, these equations
of motion are solved (with
causal boundary conditions) for some of the correlation functions, say $%
\{C_k,...C_{l+1}\}$, and the result is substituted into the $l$ loop
effective action. (We say that $\{C_k,...C_{l+1}\}$ have been {\sl slaved} to
$\{\phi ,G,C_3,...C_{k-1}\}$) The resulting truncated effective action is
generally complex and  the mean field equations of motion it generates
will come out to be dissipative, which indicates that the effective dynamics
is stochastic.

We shall stop the formal development at this point and consider the
relationship of these truncated `finite' theories and the consistent
histories approach to quantum mechanics, on which rests, in the final analysis,
the physical interpretation of any quantum formalism.
After that we will consider some familiar examples.

\section{Truncated Effective Actions and Decoherence}

The basic tenet of the consistent histories approach to quantum mechanics
\cite{conhis} is that quantum evolution may be
considered as the result of the coherent superposition of virtual fine
grained histories, each carrying full information on the state of the system
at any given time. If we adopt the `natural' procedure of specifying a
fine grained history by defining the value of the field $\Phi (x)$ at every
space time point, these field values being c numbers, then the quantum
mechanical amplitude for a given history is $\Psi [\Phi ]\sim e^{iS[\Phi ]}$,
where $S$ is the classical action evaluated at that particular history.
The virtual nature of these histories is manifested through the occurrence
of interference phenomena between pairs of histories. The strength of these
effects is measured by the `decoherence functional'

\begin{equation}
\label{popa}D_F[\Phi ,\Phi ^{\prime}]\sim \Psi [\Phi ]\Psi [\Phi
^{\prime}]^*\sim e^{i(S[\Phi ]-S[\Phi ^{\prime}])}
\end{equation}
In reality, actual observations correspond to `coarse grained' histories.
A coarse-grained history is defined in terms of a `filter function' $\alpha $,
which determines which fine grained histories belong
to the superposition, and their relative phases. For example, we may
define a coarse-grained history of a system with two degrees of
freedom $x$ and $y$ by specifying the values $x_0(t)$ of $x$ at all times.
Then the
filter function is $\alpha [x,y]=\prod_{t\in R}\delta (x(t)-x_0(t))$. The
quantum mechanical amplitude for the coarse-grained history is defined as

\begin{equation}
\label{psiofalfa}\Psi [\alpha ]=\int~D\Phi~e^{iS}\alpha [\Phi ]
\end{equation}
where the information on the quantum state of the field is assumed to have
been included in the measure and/or the boundary conditions for the
functional integral. The decoherence functional for two coarse-grained
histories is  \cite{DCH}

\begin{equation}
\label{dofalfa}D_F[\alpha, \alpha ^{\prime}]=\int~D\Phi D\Phi'
e^{i(S[\Phi]-S[\Phi'])} \alpha [\Phi ]\alpha ^{\prime}[\Phi' ]^*
\end{equation}
In this path integral expression, the two histories $\Phi$ and $\Phi'$
are not independent; they assume identical values on a $t=T={\rm constant}$
surface in the far future. These are, of course, the same boundary
conditions satisfied by the histories on each branch of the time path in the
CTPEA discussed above.

To obtain a formulation of the consistent histories approach suitable for
our present purpose, we shall (by appeal to Haag's `reconstruction theorem',
which states that the set of all expectation values of time - ordered
products of fields carries full information about the state of the
system \cite{Haag}) consider fine grained histories as specified by the
given values of the irreducible time - ordered correlation functions. These
histories include those defined by the local value of the field, as those
where all irreducible Feynman functions vanish, but allow also more
general possibilities. Coarse grained histories will be specified by finite
sets of Feynman functions, and correspond to the truncated theories
described above.

In particular, consider two histories defined by two sets of mean fields,
Feynman propagators and correlation functions up to $l$ particles, $\{\Phi
,G,C_3,...C_l\}$ and $\{\Phi ^{\prime },G^{\prime },C_3^{\prime
},...C_l^{\prime }\}$, respectively. Then the decoherence functional between
them is given by \cite{DCH}

\begin{equation}
\label{papa}D_F[\{\Phi, G, C_3,...C_l\}, \{\Phi^{\prime}, G^{\prime},
C^{\prime}_3,...C^{\prime}_l\} ]\sim e^{i\Gamma_l}
\end{equation}
where $\Gamma_l$ is the $l$-loop CTP effective action evaluated at the
following history:
(Here we have disregarded a prefactor which is not important
to our discussion.)

a) Correlation functions on the `direct' branch are defined according to
the first history: $\Phi=\phi$, $G^{11}=G$, etc.

b) Those on the `return' branch are identified with the time-reverse of those
in the second history: $\Phi'=\phi'$, $G^{22}=(G^{\prime })^{*}$, etc.

c) All others are slaved to these.

We have already discussed in detail the rationale of the ansatz (\ref{papa})
\cite{DCH}. This ansatz generalizes
the decoherence functional  (\ref{popa}) between
histories defined in the usual way, in that they reduce to it when all
irreducible Green functions are chosen to vanish, and it is consistent, in
the sense that further integration over the higher correlation functions
$\{C_{k+1},...C_l\}$, say, gives back the decoherence functional appropriate
to the $k$ loop theory in the saddle point approximation to the trace.

In the spirit of the closed-time path method we might wish to
consider more general histories, where not only time-ordered
products, but also others are freely specified. However, this cannot be done
in the present framework, because we are describing each of the two histories
entering into the decoherence functional as a c -number field distribution
defined on a simple time path. Thus, we cannot define different composite
operators. Whenever we write down a product of fields at different points,
it is automatically time ordered by the path integral. We could achieve our
goal, notwithstanding, if the fine grained histories were themselves defined
on a closed time path, thus lifting the restriction to time ordered
products. We shall not consider here these extensions of the theory.

In actual practice, we are not even interested in exhausting the possible
accuracy at a given number of loops, but rather in how further
higher correlation functions are slaved to the lower ones.
The relevant effective action under the ansatz (\ref{papa}) then
becomes the truncated effective action we discussed above.
Because the truncated effective
action has in general a positive imaginary part, this implies a suppression
of the overlap between different truncated histories, or decoherence.

Decoherence means physically that the different coarse grained histories
making up the full quantum evolution are individually realizable and may
thus be assigned definite probabilities in the classical sense.
Therefore, the quantum nature of the system will be shredded to the degree of
accuracy afforded by the coarse graining procedure, and the dynamics
be described by a  self-consistent, coupled set of equations
of a finite number of (non local) c-number variables.

In this finite, truncated theory,  decoherence is associated to information
degradation and loss of full predictability \cite{pazsinha}.
This dynamics is, however, non-deterministic and contains necessarily a
stochastic component. We shall see here that it is a consequence, within the
consistent histories language, of the categorical relationship between
dissipation and noise embodied in the fluctuation dissipation relation
\cite{fdrsc}.

At this point it is perhaps useful to look at some concrete examples.
We shall consider the stochastic dynamics of respectively the mean field
and the Feynman functions of a self interacting scalar field.

\subsection{Decoherence, Dissipation and Noise in $\lambda \Phi ^4$ theory}

Let us now look at the example of a self interacting scalar field theory
with classical action

\begin{equation}
S\left[ \Phi \right] =\int d^4x\;\left\{ - \frac{1}{2}  \left(
\partial ^\mu \Phi \partial _\mu \Phi +m^2\Phi ^2\right) -
\frac \lambda {4!} \Phi ^4\right\}
\end{equation}
and consider fluctuations of the mean field around the symmetric state.
We shall first analyze the theory at one and two loop order.
In the next section we shall then
adopt the new `correlation histories'  viewpoint  \cite{DCH},
which will require consideration of three loop effects.

\subsubsection{Dynamics of the mean field: one loop analysis}

It should be clear from the above that the one loop expansion has very special
features. To this order all $\chi$'s vanish, so only the background
field $\phi$ and the two-point functions $G$ remain, and there is no Legendre
transformation to perform. We find

\begin{equation}
\Gamma_{\infty}[\phi, G,\{C_r\}] \rightarrow \Gamma_1[\phi, G]
\equiv  S[\phi ]+ \frac{1}{2} G^{ij}\frac{\delta S[\phi ]}{\delta\phi^i
\delta \phi^j}-\frac{i}{2} \ln ~{\rm Det}~G
\end{equation}
By symmetry, there is a solution of the semiclassical equations with $\phi
\equiv 0$ and

\begin{equation}
S_{2jk}-iG^{-1}_{jk}=0
\end{equation}
Writing (signature -+++) $S_{2jk}=(\Box -m^2)g_{jk}$, and taking into
account the CTP boundary conditions, we get via

\be
G^{ab}(x,x^{\prime })=\int ~{\frac{d^dk}{(2\pi )^d}}e^{-ik(x-x^{\prime
})}G^{ab}(k)
\te
the usual result \cite{ch87}
\bea
G^{11} &=& \left( G^{22}\right) ^{*}=(-i)(k^2+m^2-i\epsilon )^{-1} \nn \\
G^{12}(k)&=& G^{21}(-k)=2\pi \theta (k^0)\delta (k^2+m^2)
\tea

We now look for the dynamics (and coherence properties) of fluctuations
around the zero point. Again, because of the field- reversal symmetry, we
know the fluctuations in the background field cannot couple to the
fluctuations in the propagators (to quadratic order). For simplicity, we
shall omit these, and consider only a perturbation $\Delta \phi $ on the
background field. The quadratic part of the effective action becomes

\be
\Delta \Gamma_1[\Delta \phi ]\equiv \ {\frac 12}\Delta \phi ^j
[(\Box-m^2)g_{jk}-\lambda _{jklm}G^{lm}]\Delta \phi ^k
\te
where
\be
\lambda_{ijkl} = \pm\lambda\delta(x_j-x_i)\delta(x_k-x_i)\delta(x_l-x_i)
\te
for $a_i=a_j=a_k=a_l=1, 2$ respectively for the plus and minus signs,
and $\lambda_{ijkl}= 0$ otherwise.
Since only the coincidence
limit of the propagators is involved, the effective action is real, and so
we reach our first conclusion: {\it There is no quantum to classical
transition for fluctuations around the symmetric vacuum at one loop order,
and, in particular, no such transition for free field theories}
\footnote{An unrelenting critic could observe that there is an imaginary
part in the one loop effective action,
since we need to add the usual $i\epsilon $'s to the mass to
enforce the vacuum initial conditions. Since the limit $\epsilon $$\to 0$
has to be taken anyway, the resulting noise is infinitely weak. It only
means that a `stable' particle with infinite lifetime is only an
idealized conception. As observed by Schwinger, any physically observable
particle has necessarily a finite lifetime \cite{ftup}}.

\subsubsection{Dynamics of the mean field: two loops analysis}

As discussed above, two loops is the first order where a nontrivial
truncation is called for. At this order, we have two graphs contributing to $%
W_2$

\be
W_2={\frac{i}{12}}(\sigma_3+\chi_3)_{ijk}G^{ii^{\prime}}G^{jj^{%
\prime}}G^{kk^{\prime}} (\sigma_3+\chi_3)_{i^{\prime}j^{\prime}k^{\prime}}+{%
\ \frac{1}{8}}(\sigma_4+\chi_4)_{ijkl}G^{ij}G^{kl}%
\te
Observe that with the present conventions, $\sigma _{4ijkl}=-\lambda _{ijkl}$,
so the second graph agrees with (3.6) in  \cite{ch88}. As discussed
above, the variation of $W_2$ with respect to $\chi _4$ yields the (time
symmetric) constraint $C_4\equiv 0$. Moreover, since $\chi _4$ enters
linearly in $W_2$, it disappears in the Legendre transform.
So the only nontrivial correlation, besides the $G$s, is $C_3$.
Now, variation with respect to $\chi _3$ yields

\be
C_3^{ijk}=iG^{ii^{\prime}}G^{jj^{\prime}}G^{kk^{\prime}}(\sigma_3+%
\chi_3)_{i^{\prime}j^{\prime}k^{\prime}}%
\te
Solving for $\chi _3$, we immediately find the two-loops effective action

\bea
\Gamma_2 & \equiv & \ S[\phi ]+{\frac 12}G^{ij}
{\frac{\delta S[\phi ]}{\delta \phi ^i\delta \phi ^j}}-
{\frac i2} \ln ~{\rm Det}~G \nn \\
&  &   +{\frac i{12}}C_3^{ijk}G_{ii^{\prime }}^{-1}G_{jj^{\prime}}^{-1}
G_{kk^{\prime }}^{-1}C_3^{i^{\prime }j^{\prime }k^{\prime }}+{\frac 16}
\sigma _{3ijk}C_3^{ijk}+{\frac 18}\sigma _{4ijkl}G^{ij}G^{kl}
\tea
As with one loop order, symmetry implies the existence of a solution with $%
\phi =C_3=0$, and

\be
S_{2jk}-iG^{-1}_{jk}-{\frac{1}{2}}\lambda_{jklm}G^{lm}=0%
\te
The modification with respect to the one loop equation amounts to the
resummation of daisy graphs. Observe that these equations remain
time-reversal invariant. Indeed, we could assume that $m^2$ is already
the physical mass, so the tadpole graph
vanishes, and the propagators would remain the same as before.

The difference between one and two loops physics appears when we look at
fluctuations. Indeed, we can still assume that the propagators remain
unperturbed; however, the perturbation $\Delta \phi ^i$ to the background
field now couples to the perturbation $\Delta C_3$ to the three point
correlation. The quadratic action now reads

\bea
\Delta \Gamma_2 & \equiv &
{\frac 12}\Delta \phi ^j[(\Box -m^2)g_{jk}-\lambda _{jklm}G^{lm}]\Delta
\phi ^k \nn \\
&  &   +{\frac i{12}}\Delta C_3^{ijk}G_{ii^{\prime }}^{-1}G_{jj^{\prime}}^{-1}
G_{kk^{\prime }}^{-1}\Delta C_3^{i^{\prime }j^{\prime }k^{\prime }}
-{\frac 16}\lambda _{ijkl}\Delta \Phi ^l\Delta C_3^{ijk}
\tea
Still there is no sign of dissipation or loss of coherence. However, we do
get the nontrivial `mean field' equation

\be
iG^{-1}_{ii^{\prime}}G^{-1}_{jj^{\prime}}G^{-1}_{kk^{\prime}} \Delta
C_3^{i^{\prime}j^{\prime}k^{\prime}}-\lambda_{ijkl}\Delta\phi^l=0
\te
for the three point function. If one imposes the time-oriented ansatz

\be
\Delta C_3^{ijk}=-iG^{ii^{\prime}}G^{jj^{\prime}}G^{kk^{\prime}}
\lambda_{i^{\prime}j^{\prime}k^{\prime}l^{\prime}}\Delta\phi^{l^{\prime}}
\te
and substitute into the effective action, we obtain the truncated effective
action

\be
\Delta \Gamma _{2T}[\Delta \phi ]\equiv \ {\frac 12}\Delta \phi ^j((\Box
-m^2)g_{jk}-\lambda _{jklm}G^{lm})\Delta \phi ^k+{\frac i{12}}\Delta
\phi^j\Sigma _{jk}\Delta \phi ^k
\te
where $\Sigma_{jk}$ represents the setting sun graph

$$
\Sigma_{ll^{\prime}}=\lambda_{ijkl}G^{ii^{\prime}}G^{jj^{\prime}}G^{kk^{%
\prime}}\lambda_{i^{\prime}j^{\prime}k^{\prime}l^{\prime}}%
$$

Our ansatz represents, of course, the slaving of the fluctuations in the
correlations to those in the background field. Moreover, in the physical
limit where $\phi=\phi'$, the relationship is causal. In other words,
we ascribe any deviation from Gaussian correlations to the presence of
quanta emitted by the fluctuating background field.

The new term in the effective action can be written in the form (\ref{R14})
\cite{nfsg}, where

\be
{\cal D}(x,x^{\prime })= [-\Box +m^2+\lambda G_F(x,x)]\delta (x-x^{\prime})
+D(x,x^{\prime }),
\te
where
\bea
D(x,x^{\prime})&=&{\frac{\lambda^2}{3}}[{\rm Im}(G^{11}(x,x^{\prime}))^3]
\theta (t-t^{\prime}) \nn \\
N(x,x^{\prime})&=&{\frac{\lambda^2}{6}}[{\rm Re}(G^{11}(x,x^{\prime}))^3]
\tea
which makes manifest the real and imaginary parts. The relevant graph has
been computed in detail in  \cite{ch88} (Eq. 4.27). However,
there is a more direct
way to appreciate the physical meaning of these kernels.
 From Poincar\'e invariance and the time - ordered property we must have

\be
[G^{11}(x,x^{\prime })]^3={\frac{-i\mu ^{2\epsilon }}{(4\pi )^4}}\int ~{\
\frac{d^dk}{(2\pi )^d}}~e^{-ik(x-x^{\prime })}F(-k^2+i\epsilon )
\te
where $F$ itself admits the Lehmann representation \cite{bjorkendrell}

\be
F(z)=a+b(z-m^2)+(z-m^2)^2\int_{9m^2}^{\infty}{\frac{ds~h(s)}{(s-z)(s-m^2)^2}}%
\te
where $a$ and $b$ are actually divergent
\bea
a &=& 3m^2[{\frac{2}{\epsilon^2}}+{\frac{2}{\epsilon}}(A+B\ln {\frac{m^2}
{4\pi\mu^2}}) + \ln^2 {\frac{m^2}{4\pi\mu^2}}+C\ln {\frac{m^2}{4\pi\mu^2}}+D]
\nn \\
b &=&   {\frac{1}{2\epsilon}}+{\frac{1}{2}}\ln {\frac{m^2}{4\pi\mu^2}}+E
\tea
Here, $A,B,C,D,E$ are numerical constants, and $\epsilon =d-4$  \cite{ch88}.
These terms may be absorbed into the classical action and need not worry us.
More relevant to our concern is the $h$ function, which gives the
imaginary part of $F$. It must be positive, because of the optical theorem.
An actual evaluation yields

\be
h(s)={\frac{16}s}\int_{4m^2}^{(\sqrt{s}-m)^2}dt\sqrt{(s+m^2-t)^2-4sm^2}\sqrt{%
1-{\frac{4m^2}t}}\theta \left( s-9m^2\right)
\te
With this knowledge, it is immediate to find the Fourier expansions

\bea
D(x,x^{\prime }) &=& {\frac{\lambda ^2}6}{\frac{\mu ^{2\epsilon }}{(4\pi )^4}}
\int ~{\frac{d^dk}{(2\pi )^d}}~e^{-ik(x-x^{\prime })}\int_{9m^2}^\infty
{\frac{ds~h(s)}{(s+(k+i\epsilon )^2)}} \nn \\
N(x,x^{\prime}) &=&  {\frac{\lambda^2}{6}}{\frac{\mu^{2\epsilon}}{(4\pi )^4}}
\int~{\frac{d^dk}{(2\pi )^d}}~e^{-ik(x-x^{\prime})}\pi h(-k^2)\theta
(-k^2-9m^2)
\tea
where $(k+i\epsilon )^2=-(k_0+i\epsilon )^2+\vec k^2$. Associating as usual
the imaginary part of the effective action with noise and decoherence, and
the imaginary part (in momentum space) of the retarded propagator with
dissipation, we conclude at once that: {\it Only the amplitudes
corresponding to momenta above the three particle threshold lose their
quantum coherence and become classical. Therefore, only those
amplitudes are subject to random fluctuations in the semiclassical regime.}
The corresponding statement in terms of dissipation would be:
{\it Only the
amplitudes corresponding to momenta above the three particle threshold are
subject to dissipation} .

The presence of a fluctuation - dissipation relation underlying all this
should be obvious, but we can be more explicit still. We know that the noise
affecting the above- threshold amplitudes will have a mean square value

\be
\sigma ^2(k)\sim N(k)\sim {\frac{\lambda ^2}6}{\frac{\mu ^{2\epsilon }}
{(4\pi )^4}}\pi h(-k^2)
\te
The FDR is then just the statement that

\be
{\rm Im}~G_{ret}\sim |G_{ret}|^2N(k)~~{\rm when} -k^2>9m^2,
\te
thus linking the noise autocorrelation to the
spectrum of dissipated energy, just as in the early formulations by Callen
and Welton and as we have illustrated in the cosmological particle creation
problem \cite{ch87,ch89}.
Callen and Welton's formulation of the theorem is recovered by identifying

\be
Z=\left( i\omega G_{ret}\right) ^{-1} , ~~~
R={\rm Re}Z=\left( \frac 1\omega \right) N(k)
\te
So, given $\vec k$, we find
\be
\sigma ^2=\int_{-\infty }^\infty \frac{d\omega }{2\pi }N(k)
= \frac 2\pi \int_0^\infty d\omega \left( \frac \omega 2\right) R\left(
\omega \right)
\te
as expected.

This leads  to an important point, namely that one should not be too quick
in identifying the Hadamard function of the quantum field with the
correlator of the stochastic field. In our case, the quantum Hadamard
function also has an on-shell contribution; however, this part does not
appear in the classical correlator, as on shell fluctuations are dissipation
free, and thus never become classical. Another popular application
is galaxy formation seeded by quantum fluctuations in inflationary cosmology.
The conventional wisdom is to identify straightforwardly  the classical
correlator with the full Hadamard kernel, rather than only to the part
partaking in decoherence, as we assert here. The conventional way leads
to a gross overestimation of the amplitude of the resulting inhomogeneities
\cite{fgic}.

This distinction between on and off shell fluctuations disappears in the
important case of massless fields. In this case, we get $h(-k^2)\sim -8k^2$
for every time-like 4-momentum. Still, different components of the field
will have different characteristic times at which they become classical.

\section{Correlation Noise}

While describing the dynamics of a quantum field in terms of the mean field
or order parameter is the most familiar approach to causal field theory, in
many applications the relevant information is most readily retrieved not
from the mean field, but from the correlation functions \cite{sdqft}.
Moreover, as we have seen above, the sequence of correlations introduces in
non linear field theory an intrinsic hierarchical ordering, which allows us
to fill the interpolating steps between the complete unitary underlying
field theory and the truncated effective descriptions that account for our
actual experience, without imposing any {\sl ad hoc} concept of relevance
or distinction from outside.
As a demonstration of this more sophisticated approach to
decoherence and noise in field theory, in this section we shall continue
to work with the $\lambda \Phi ^4$ theory as example. Here we shall
simply assume that the mean field vanishes identically (as it usually does,
because of symmetry) and focus directly on the dynamics of the two point
functions.

\subsection{Naive derivation of `correlation noise'}

As it turns out, the case in point is simple enough to
deduce the Langevin type of Schwinger-Dyson equation for the  propagators
by an {\sl ad hoc} argument, totally independent of the MEA formalism.
It is instructive to review this argument, since it sheds light on the
physics involved, and will serve as a check on our formulation
when we proceed with the formal development below.

We start with  the Heisenberg equation of motion for the field operator

\begin{equation}
\label{heiseneq}\left( \Box -m^2\right) \Phi \left( x\right) -
\frac\lambda 6 \Phi ^3\left( x\right) =0
\end{equation}
and the canonical commutation relation

\begin{equation}
\label{ccr}\left[ \frac{\partial \Phi }{\partial t}\left( t,\vec x_1\right)
,\Phi \left( t,\vec x_2\right) \right] = -i\delta (\vec x_1-\vec x_2)
\end{equation}
Let us introduce the time ordered product of fields
\begin{equation}
\label{top}T\left[ \Phi \left( x_1\right) \Phi \left( x_2\right) \right]
=\theta \left( t_1-t_2\right) \Phi \left( x_1\right) \Phi \left( x_2\right)
+\theta \left( t_2-t_1\right) \Phi \left( x_2\right) \Phi \left( x_1\right)
\end{equation}
 From ( \ref{heiseneq}) and (\ref{ccr}) we derive the operator equation

\begin{equation}
\label{opeqmot}\left( \Box _1-m^2\right) T\left[ \Phi \left( x_1\right) \Phi
\left( x_2\right) \right] -\frac \lambda 6 T\left[ \Phi
^3\left( x_1\right) \Phi \left( x_2\right) \right] =i\delta \left(
x_1-x_2\right)
\end{equation}

In the present context, a correlation history is defined by selecting those
field histories where $T\left[ \Phi ^i\Phi ^j\right] $ remains `close'
to a given kernel $G_F$. If Wick's theorem holds, then we could write also
the second term in  (\ref{opeqmot} ) in terms of binary field products, as

\bea
T\left[ \Phi ^3\left( x_1\right) \Phi \left( x_2\right) \right] &\sim &
3G_F\left( x_1,x_1\right) G_F\left( x_1,x_2\right) \nn \\
&- & i\lambda \int d^4y [ G_F^3\left( x_1,y\right) G_F\left( y,x_2\right)
-G^{-3}\left( x_1,y\right) G^{-}\left( y,x_2\right) ] +\ldots
\tea
Here we are not writing explicitly the one- particle or higher order
reducible graphs,  $G^{-}$ stands for the  `value ' of the
composite operator $\Phi (x_1)\Phi (x_2)$ in the given correlation history.
Since Wick's formula is not exact, however, the best we can do is to write

\bea
&& \left( \Box _1-m^2 \right) G_F \left( x_1,x_2 \right) - \frac{1}{2} \lambda
G_F \left( x_1,x_1 \right) G_F \left( x_1,x_2 \right) \nn \\
&+& \frac{i}{6}\lambda ^2 \int d^4y [G_F^3 \left( x_1,y \right)
G_F \left( y,x_2 \right) -G^{-3} \left( x_1,y \right) G^{-}\left( y,x_2
\right)]
+\ldots \nn \\
&= & i\delta \left( x_1-x_2 \right)+ \frac{1}{2} \lambda f \left(x_1,x_2
\right)
\tea
where

\be
f\left( x_1,x_2\right) \sim  \frac 13 T\left[ \Phi ^3\left(
x_1\right) \Phi \left( x_2\right) \right] -G_F\left( x_1,x_1\right)
G_F\left( x_1,x_2\right)
\te
is  `small '. The equation for $G_F$ simplifies substantially if we
operate on it from the left with $G_F^{-1}\sim -i(\Box -m^2)$. We get

\bea
\label{lsdeq2}
-iG_F^{-1}\left( x_1,x_2\right) +\left( \Box -m^2\right) \delta \left(
x_1,x_2\right)
&-& \frac {1}{2} \lambda  G_F\left( x_1,x_1\right) \delta \left(
x_1,x_2\right) + \frac{i}{6} \lambda ^2 G_F^3\left( x_1,x_2\right) \nn \\
& = & \frac {1}{2} \lambda  K\left( x_1,x_2\right) \delta \left(
x_1,x_2\right) ,
\tea
where now all the terms up to order $\lambda ^2$ are explicit, and the
driving force is

\begin{equation}
\label{naivestf}K\left( x_1,x_2\right) \sim \left\{ \Phi \left( x_1\right)
\Phi \left( x_2\right) -G_F\left( x_1,x_2\right) \right\}.
\end{equation}

Of course, $K$ is a q-number quantity , which reminds us that $G_F$ is a
composite operator. However, after decoherence, we can look upon $G_F$
as the actual value of the time ordered product {\sl and} a c-number quantity.
In the semiclassical regime, $K$ becomes a {\sl bona fide} classical stochastic
force. In a first approximation, we can take this force to be Gaussian,
with a zero mean. Following the usual procedure  \cite{Landau},
we derive its standard deviation from the symmetric expectation
value

\be
\left\langle K\left( x_1,x_2\right) K^{*}\left( y_1,y_2\right) \right\rangle
_c\equiv  \frac 12 \left\langle \left\{ K\left( x_1,x_2\right)
,K^{\dagger} \left( y_1,y_2\right) \right\} \right\rangle _q
\te
to be

\begin{equation}
\label{noisecorr}
\left\langle K\left( x_1,x_2\right) K^{*}\left( y_1,y_2\right) \right\rangle_c
= \frac 12 \left\{ G^{+}\left( x_1,y_1\right) G^{+}\left(
x_2,y_2\right) +G^{-}\left( x_1,y_1\right) G^{-}\left( x_2,y_2\right)
+\left( y_1\rightleftharpoons y_2\right) \right\}
\end{equation}

The  `naive' derivation of the Langevin type Schwinger - Dyson equation
may be carried through directly at the level of fields defined on a closed
time path. We only quote the result

\begin{equation}
\label{lanctpeq}
 -\frac{i}{2} \left( G^{-1}\right) _{ii^{\prime }}+{\frac 12}
 g_{ii^{\prime }}(\Box -m^2)- \frac 14 \lambda _{ii^{\prime}
 jj^{\prime }}G^{jj^{\prime }}
+ \frac i{12} \lambda _{ijkl}G^{jj^{\prime }}G^{kk^{\prime
}}G^{ll^{\prime }}\lambda _{i^{\prime }j^{\prime }k^{\prime }l^{\prime }}
= \frac 12 \lambda _{ii^{\prime }jj^{\prime }}K^{jj^{\prime }}
\end{equation}
where
\be
K^{ij}\sim \Phi ^i\Phi ^j-G^{ij}
\te
Upon decoherence $K$ becomes a classical stochastic source. Observe that the
(12) and (21) equations are not modified; also that $K$ is generally a
complex source, with $K^{22}=(K^{11})^{*}$. We recover a positive definite
noise autocorrelator as

\be
\left\langle K^{11}K^{22}\right\rangle _c=\left\langle K\left(
x_1,x_2\right) K^{*}\left( y_1,y_2\right) \right\rangle _c
\te
The right hand side is given, again, by (\ref{noisecorr}). This is
the `naive' derivation of the Langevin equation for the
Feynman propagator.

\subsection{Systematic study of `correlation noise' }

We shall now show how to recover the results of our `naive' approach
to the stochastic behavior of Green functions, by systematically following
the theory sketched in the previous Sections.

We need to carry out our analysis at three loops order, this
being the lowest order at which the dynamics of the correlations is
nontrivial, in the absence of a symmetry breaking background field \cite{ch88}.
To this accuracy, we have room for four nonlocal sources besides the mean
field and the two point correlations, namely $\chi _3$, $\chi _4$,$\chi _5$
, and $\chi _6$. However, the last two enter linearly in the generating
functional. Thus the three- loop effective action only depends nontrivially
on the mean field and the two, three and four point correlations. By
symmetry, there must be a solution where the mean field and the three point
function remain identically zero, which we shall assume. Slaving the four
point function to the propagators, we find the truncated action \cite{ch88}

\be
\Gamma _3[G]  \equiv {\frac 12}g_{ij}(\Box -m^2)G^{ij}-{\frac i2}
\ln ~{\rm Det}~G - \frac 18 \lambda _{ii^{\prime }jj^{\prime}}
G^{ii^{\prime }}G^{jj^{\prime }}+ \frac i{48} \lambda_{ijkl}
G^{ii^{\prime }}G^{jj^{\prime }}G^{kk^{\prime }}G^{ll^{\prime}}
\lambda _{i^{\prime }j^{\prime }k^{\prime }l^{\prime }}
\te
and the mean field equation of motion for the propagators

\be
- \frac{i}{2} \left( G^{-1}\right) _{ii^{\prime }}+{\frac 12}
 g_{ii^{\prime }}(\Box -m^2)-\frac 14 \lambda _{ii^{\prime}j
 j^{\prime }}G^{jj^{\prime }}+ \frac i{12}
 \lambda_{ijkl}G^{jj^{\prime }}G^{kk^{\prime }}G^{ll^{\prime }}
 \lambda _{i^{\prime}j^{\prime }k^{\prime }l^{\prime }}=0
\te
which reduces to the left hand side of (\ref{lanctpeq}) .

To derive the right hand side of the Langevin type Dyson-Schwinger
equation, we must consider fluctuations around the mean field solution.
Concretely, we expect to retrieve the noise autocorrelation (\ref{naivestf})
from the second variation of the two particle irreducible (2PI)
effective action, evaluated at the extremum point.
A perturbation $\Delta ^{ij}$ to the propagators $G^{ij}$ will satisfy the
equations

\begin{equation}
\label{lpf} \frac i2 \left( G^{-1}\right) _{ij}\Delta^{jj^{\prime }}
\left( G^{-1}\right) _{j^{\prime }i^{\prime }}-\frac14
\lambda_{ii^{\prime }jj^{\prime }}\Delta ^{jj^{\prime }}+\frac i4
\lambda _{ijkl}G^{jj^{\prime }}G^{kk^{\prime }}\Delta^{ll^{\prime }}
\lambda _{i^{\prime }j^{\prime }k^{\prime }l^{\prime }}=0
\end{equation}
The linearized equations for the perturbations follow from the quadratic 2PI
effective action  \cite{2pi}  \cite{ch88}

\begin{equation}
\label{lin2piea}
\Delta \Gamma =   \frac i4 {\rm Tr}\left[ \left( G^{-1}\right)_{ij}
\Delta^{ji^{\prime }}\right] ^2
- \frac 18 \lambda _{ii^{\prime }jj^{\prime }}
\Delta^{ii^{\prime }}\Delta ^{jj^{\prime }}
+\frac i8 \lambda _{ijkl}\Delta ^{ii^{\prime }}G^{jj^{\prime}}
G^{kk^{\prime }}\Delta ^{ll^{\prime }}\lambda _{i^{\prime }j^{\prime}
k^{\prime }l^{\prime }}
\end{equation}

The connection between the effective action and the noise follows from the
observation that the decoherence functional between a correlation history
with the Feynman propagator, say, $G_F+\Delta ^{11}$, and another
with $G_F+(\Delta ^{22})^{*}$ should be given by $D_F \sim {\rm \exp }
(i\Delta \Gamma )$. However, in this case we cannot directly identify
$\Delta \Gamma $ from (\ref{lin2piea}) as the  `phase' of the
decoherence functional. This is obvious already from the fact that the
decoherence functional depends only on two kernels $\Delta ^{11}$ and $%
(\Delta ^{22})^{*}$, while $\Delta \Gamma $ depends on four kernels, both
the above and also the positive and negative frequency propagators $\Delta
^{21}$ and $\Delta ^{12}$. In order to retrieve the decoherence functional
from the quadratic effective action,  we must first  `slave '
these excess kernels to the time ordered ones. We do this by extremizing
$\Delta \Gamma $ with respect to $\Delta ^{12,21}$, holding $\Delta ^{11,22}$
fixed. Only when we substitute the slaved propagators back into the
effective action can we establish the connection to decoherence and noise.

Because of the inherent complexity of the 2PI effective
action, it is necessary to impose certain restrictions on the variations
allowed on the propagators, in order to arrive at meaningful results. We
shall consider only real perturbations of the propagators, and assume that
the linearized equations hold at $O\left( \lambda \right) $
(Note that in  \cite{DCH} a different set of restrictions were imposed, so
the results below are correspondingly different.)

As in the previous section, it is convenient to switch to variables
describing the propagators on the ` `diagonal' ' of the decoherence
functional, that is, when both histories are the same, and the departure
from this diagonal. Since on the diagonal we have $\Delta ^{11}=\Delta ^{22}$
it is convenient to define

\be
\left\{ \Delta \right\} =\Delta ^{11}+\Delta ^{22} ,~~
\left[ \Delta \right] =\Delta ^{11}-\Delta ^{22}
\te

%
%
%
%
%

According to the CTP boundary conditions, $\Delta^{12}(x,x')$ should turn into
the Dyson function as $x$ ''goes round the corner'' at $T\to\infty$,
and similarly $\Delta^{21}$ turns into the Feynman function. This dictates the
boundary conditions at infinity, and since all propagators involved are Klein
Gordon solutions at tree level, this means that

\be
\Delta^{21}(x,x')\equiv\Delta^{11}(x,x'); ~~
\Delta^{12}(x,x')\equiv\Delta^{22}(x,x')
\te
throughout, at tree level. The $O(\lambda^2)$ terms
which have the generic form
\bea
i\lambda^2\int~d^4xd^4x'& &\{ G_F^2(x,x')(\Delta^{11})^2(x,x')
-G_+^2(x,x')(\Delta^{21})^2(x,x') \nn \\
&& -G_-^2(x,x')(\Delta^{12})^2(x,x')
+G_D^2(x,x')(\Delta^{22})^2(x,x')\}
\tea
now become

\be
i\lambda^2\int~d^4xd^4x'~\{\theta(t'-t)[G_-^2(x,x')-G_+^2(x,x')]
                                     [(\Delta^{11})^2-(\Delta^{22})^2](x,x')\}
\te
They are manifestly real, and are thus unrelated to noise.


Finally, observe that the linearized equations imply, at $O(\lambda )$, the
identity

\be
\left[ G^{-1}\Delta \right]_i^j \left( x,y \right) =- \frac{i}{2}\lambda
\left(
\begin{array}{ccc}
\Delta ^{11}\left( x,x\right) G^{11}\left( x,y\right)
&  & \Delta^{11}\left( x,x\right) G^{12}\left( x,y\right) \\
&  &  \\
-\Delta ^{22}\left( x,x\right) G^{21}\left( x,y\right)
&  & -\Delta^{22}\left( x,x\right) G^{22}\left( x,y \right)
\end{array}
\right)
\te
Computing the trace of the square of this matrix, and substituting into the
linearized effective action, we recover its imaginary part as

\be
{\rm Im}\left[ \Delta \Gamma \right] =- \frac{1}{32}\lambda^2
\int d^4xd^4y\left\{ \left[ \Delta \right] ( x,x) \left[ \left(
G^{+}( x,y) \right) ^2+\left( G^{-}( x,y) \right)^2 \right] \right\}
\left[ \Delta \right] \left( y,y\right)
\te
The  `wrong' sign reminds us that the stochastic forces are generally
complex. However, the correlator

\be
\left\langle KK^{*}\right\rangle \sim -\left( \frac{\partial ^2}{\partial
\left[ \Delta \right] ^2}\right) {\rm Im}\left[ \Delta \Gamma \right]
\te
comes out positive, as it should. This gives the same result as from our
earlier `naive' derivation.

{}.

\section{Discussions}

In this paper  we  have presented a new approach to the analysis of the
statistical mechanics of interacting  quantum  field  theory
based on the properties and dynamics of the infinite number of
correlation functions of the field. We point out that ordinary perturbative
field theory involving only a finite number of correlation functions
amounts to an open system obtained by truncating the
Dyson- Schwinger equations for the hierarchy  of  correlations.
Inclusion of the effect of the higher correlation functions
leads to modifications of the dynamics of the lower order functions
appearing  as fluctuations, noise, dissipation and irreversibility.
This proves our earlier conjecture \cite{ch88,HuPhysica,Banff},
that any effective field theory is intrinsically dissipative and stochastic
in nature.

Our approach is inspired by the Boltzmann-BBGKY  formulation  of  the
kinetic theory of  dilute  gases   \cite{Akhiezer,Balescu},  and
incoporates the ideas of noise and fluctuations in the Langevin-Fokker-Planck
approach to the stochastic mechanics of Brownian motion \cite{vanKampen}.
The techniques we used are the closed time path \cite{ctp} and the multiple
source effective actions \cite{2pi}. Decoherence and quantum to
classical transition are discussed in the consistent history conceptual
framework.
We see once again \cite{HuTsukuba,GelHar2} that decoherence, dissipation and
noise  are  indissolubly  related  to  each  other  in any nontrivial
interacting theory.   Therefore  the  best classical description of the
dynamics is both irreversible  and noisy.  If the description is framed
in terms of correlations rather  than  the  more  familiar mean fields,
then the propagators themselves will become  stochastic, generating
the so called `correlation noises'.

Possible applications of the viewpoint expounded and the results derived
in this  paper are manifold,  as statistical ideas and measures such as
extraction of relevant information, formulation in terms of collective
variables, truncation and coarse graining  of irrevelant variables
permeate almost all areas of theoretical physics where one wants to derive
the macroscopic features of a system in terms of microscopic laws or when
one attempts to give a simple description to complex phenomena.
As guides to our present investigation, in addition to the paradigmatic
schemes of non-equilibrium statistical mechanics and the philosophical
underpinnings of decoherent histories approach to quantum mechanics,
it is worth mentioning that as concrete problems with deep meanings,
{\it critical dynamics} provides not only a physical model for our thoughts
but also a challenge to solving some of its outstanding problems.
For example,  there is at present no satisfactory
understanding of nonequilibrium first order phase transitions, as most work
on the subject rests on Euclidean methods which presuppose equilibrium
conditions. Following the pioneering work of Calzetta \cite{sdqft}
there have been more recent work to apply the CTP method
to the study of phase transitions in the early universe
\cite{firstorder}. Our present formalism can provide a more rigorous foundation
for these investigations.

Fluctuation phenomena giving rise to instability
of structure and growth \cite{flucgrow,HMLA} is a new and interesting
direction of potential development and application.
In this connection, an active  area  where our approach will be directly
applicable is {\it structure formation} from primordial fluctuations
in the early universe.
It is believed that these (classical) primordial fluctuations
evolved out of quantum fluctuations of the matter fields, got amplified
in the inflationary stage of the universe \cite{inflation,starobinsky},
and became classical as they grew larger than the Hubble radius.
Most existing treatments pay no attention to the subtle and important
issue of quantum to classical transition, i.e., transition from a
pure quantum to a stochastic classical description of the fluctuations.
These unsatisfactory aspects of conventional wisdom have been pointed out
by Hu, Paz and Zhang \cite{HuBelgium},
who outlined a stochastic field theoretical
derivation of the noise and fluctuations from two interacting fields.
The present authors \cite{nfsg} have also pointed out a fundamental flaw
in the conventional treatment of classical fluctuations, vis., in simply
identifying them with the Hadamard functions.
Detailed discussions of this issue from the perspective of this paper
can be found in  \cite{fgic}.

We have already mentioned the direct implication of our viewpoint and
results on {\it effective field theory} \cite{Banff}.
How successful a given theory can describe
the physical world really depends on the relevant (energy, space, or time)
scales in question and the degree of precision (or power of resolution) of
measurements. A theory can  be `perfect' at a low energy scale but makes
no sense at all at a higher energy scale. How a fundamental theory
appears at low energies or long wavelengths (infrared behavior) is a question
often asked, from gauge hierarchy and dimensional reduction in particle physics
to the approach to critical point in phase transitions. How can one
infer the attributes of the `fundamental' theory at high energies from
our limited knowledge at low energies? This is admittedly a more difficult
question, but not exactly a futile one. It has been the route taken in
our search from atomic to nuclear to particle physics, and today, towards
unified theories including quantum gravity. Understanding the nature of
effective theories is, in our opinion, essential for grasping the
interconnection of physical phenomena at different scales and interaction
ranges, and for probing into uncharted domains.
For example, it has been speculated that near the Planck time,
spacetime fluctuations will grow and the smooth manifold
picture in the semiclassical approximation will give way to stochastic
behavior before the quantum gravity regime fully takes over
\cite{nfsg,HM3,fdrsc}. The Langevin- type equations for
lower correlation functions
driven by the correlation noise in our framework can be used as models to
examine possible {\it phase transition behavior from the semiclassical to the
quantum regime} at the Planck time.

The other long-standing motivation for us to work out the statistical
mechanical
properties of quantum fields is to analyze the {\it black hole entropy and
information loss problem}. We feel that our
current understanding  of the black hole backreaction and  entropy problems
based on equilibrium thermodynamics is  inadequate. To tackle
the collapse and backreaction problems requires a knowledge of
nonequilibrium statistical mechanics for spacetime and quantum field dynamics.
We also believe that {\it nonlocality} plays an important role in the
entropy and information problem. It was for these reasons that we embarked
on developing a formalism for treating nonequilibrium fields \cite{ch88}
in curved spacetime, searching for a way to define the entropy of quantum
fields \cite{HuPavon,HuKandrup}, as well as seeking a physical understanding
of Hawking radiation in terms of scale transformations
\cite{HuEdmonton,cgea,jr}. The information flow
in an interacting quantum  field or between spacetime and matter fields
cannot be fully captured by local observations. The correlation hierarchy
scheme can incoporate and systemize the information content of the field and
its dynamics more accurately. We are now applying it to some simple
particle-field
models to understand the physics \cite{CHPR}, with the aim of treating
the corresponding problems for black holes.

At a  deeper theoretical level,  our program of research
has been to add a statistical mechanical  dimension for
quantum field  theory.  Approaching a problem treated ordinarily
with quantum field theoretical emphasis with  statistical mechanics
viewpoints can bring in refreshingly new and valuable physical insights.
Likewise, the many powerful conceptual
and computational  tools of field theory can enrich greatly both the
theoretical and applied problems of  equilibrium  and  nonequilibrium
statistical  physics. This effort has proven to be extremely
fruitful in many areas of physics.  Quantum gravity, early universe and black
hole physics are arenas where the laws of physics are pushed to the limit.
Taking up such a challenge for these problems will be even more stimulating
and rewarding.

\vskip 1cm

\noindent {\bf Acknowledgments}
We wish to thank the organizers, Steve Fulling, Gabor Kunstatter and
Tom Osborn for their kind invitation to speak at this meeting, and the
participants for their interest and discussions on our work, especially
A. Barvinsky, V. Frolov, D. Page, R. Sorkin, W. Unruh and G. Vilkovisky.
EC wishes to thank G. Gonzalez for insightful comments on the nature of
fluctuation - dissipation relations.
This work is supported in part by NSF(US) grant PHY91-19726, INT91-xxx
and by CONICET,UBA and Fundaci\'on Antorchas (Argentina).
Part of the work was done when BLH visited the Newton Institute for
Mathematical
Sciences of the University of Cambridge during the Geometry and Gravity program
in Spring 1994 and the Institute for Advanced Study, Princeton,
supported by  the Dyson Visiting Professor Fund. He also gratefully
acknowledges the support of the General Research Board of the Graduate School
of the University of Maryland for a research leave award.

\end{document}